\documentclass[12pt]{article}

\usepackage{amsmath,amssymb}
\usepackage{amsfonts}

\newcommand{\be}{\begin{equation}}
\newcommand{\ee}{\end{equation}}
\newcommand{\bea}{\begin{eqnarray}}
\newcommand{\eea}{\end{eqnarray}}
\newcommand{\ba}{\begin{array}}
\newcommand{\ea}{\end{array}}

\newcommand{\eref}[1]{(\ref{#1})}

\newcommand{\CN}{{\cal N}}

\newcommand{\BC}{{\mathbb C}}
\newcommand{\CP}[1]{{\mathbb P}^{#1}}

\newcommand{\bra}[1]{\langle{#1}|}
\newcommand{\ket}[1]{|{#1}\rangle}
\newcommand{\ip}[2]{\langle{#1}|{#2}\rangle}

\begin{document}

\rightline{hep-th/0605105}
\rightline{DCTP-06/09}
\rightline{VPI-IPPAP-06-07}

\vskip 0.75 cm
\renewcommand{\thefootnote}{\fnsymbol{footnote}}
\centerline{\Large \bf Why there is something so close to nothing:}
\centerline{\large \bf towards a fundamental theory of the cosmological constant}
\vskip 0.75 cm

\centerline{{\bf
Vishnu Jejjala${}^{1}$\footnote{\tt vishnu.jejjala@durham.ac.uk}
and
Djordje Minic${}^{2}$\footnote{\tt dminic@vt.edu}
}}
\vskip .5cm
\centerline{${}^1$\it Department of Mathematical Sciences,}
\centerline{\it Durham University,}
\centerline{\it South Road, Durham DH1 3LE, U.K.}
\vskip .5cm
\centerline{${}^2$\it Institute for Particle Physics and Astrophysics,}
\centerline{\it Department of Physics, Virginia Tech,}
\centerline{\it Blacksburg, VA 24061, U.S.A.}
\vskip .5cm

\setcounter{footnote}{0}
\renewcommand{\thefootnote}{\arabic{footnote}}

\begin{abstract}
The cosmological constant problem is turned around to argue for a new
foundational physics postulate underlying a consistent quantum theory of gravity and matter,
such as string theory. This postulate is a quantum equivalence principle which demands a consistent gauging of the
geometric structure of canonical quantum theory.
We argue that string theory can be formulated to accommodate such a principle, and that
in such a theory the observed cosmological constant is a fluctuation about a zero value.
This fluctuation arises from an uncertainty relation involving the cosmological constant and the effective volume of spacetime.
The measured, small vacuum energy is dynamically tied to the large ``size'' of the universe, thus violating
na\"{\i}ve decoupling between small and large scales.
The numerical value is related to the scale of cosmological supersymmetry breaking, supersymmetry being
needed for a non-perturbative stability of local Minkowski spacetime regions in the classical regime.
\end{abstract}

\newpage

\section{Vacuum energy in quantum gravity}
Recent cosmological observations suggest that we live in an accelerating Universe \cite{sn, wmap3}.
One possible engine for late time acceleration is an unseen ``dark energy'' that comprises 74\% of the total energy density in the Universe.
The leading candidate for dark energy is the energy in the vacuum itself, and the data suggest a small, positive cosmological constant.
This leads to a two-fold cosmological constant problem \cite{weinb}:
\begin{enumerate}
\item Why is the energy density in the vacuum so small compared to the expectation of effective field theory?
\item Why are the energy densities of vacuum and matter comparable in the present epoch?
\end{enumerate}

The first cosmological constant problem concerns both ultraviolet and infrared physics.
In quantum field theory, the cosmological constant counts the degrees of freedom in the vacuum.
Heuristically, we sum the zero-point energies of harmonic oscillators and write
\be
E_{\rm vac} = \sum_{\vec{k}} \left( \frac{1}{2}\hbar \omega_{\vec{k}} \right).
\ee
The sum is manifestly divergent.
Because quantum field theories are effective descriptions of Nature, we expect their validity to break down beyond a certain regime and be subsumed by more fundamental physics.
We may introduce a high-energy cutoff to regulate the sum, but $E_{\rm vac}$ will then scale with the cutoff.
The natural cutoff to impose on a quantum theory of gravity is the Planck energy $M_{\rm Pl}$.
This prescription yields an ultraviolet enumeration of the zero-point energy.

In the infrared, the cosmological constant feeds into Einstein's equations for gravity:
\be
R_{\mu \nu} - \frac{1}{2} g_{\mu \nu} R = 8\pi G_N \left( - \Lambda g_{\mu \nu} + T_{\mu \nu} \right).
\label{eq:einstein}
\ee
Present theories of quantum gravity are unable to deal with the cosmological constant problem.
Here we are concentrating on string theory, as the only known example of a consistent theory of
perturbative quantum gravity and Standard Model like matter.
In perturbative string theory, the dynamics of the background spacetime are determined by the vanishing of the $\beta$-functional associated to the Weyl invariance of the worldsheet quantum theory;
eq.\ (\ref{eq:einstein}) is then corrected in an $\alpha'$ expansion.
In either case, the vacuum is just any solution to these equations.
We compute the vacuum energy in quantum field theory, include it on the right hand side of the gravitational field equation, and find that spacetime is not Minkowski.
Although the Einstein equations are {\em local} differential equations, the cosmological constant sets a {\em global} scale and determines the overall dynamics of spacetime.
Our Universe is approximately four-dimensional de Sitter space (dS) with $\Lambda \approx 10^{-120} M_{\rm Pl}^4$.

Quantum theory (as presently understood) therefore grossly overcounts the number of vacuum degrees of freedom.
There is no obvious way to reconcile the generic prediction of effective field theory that the vacuum energy density should be $M_{\rm Pl}^4$ with the empirical observation that it spectacularly is not.\footnote{
There is possibly an illuminating analogy to be drawn with the problem of specific heats in pre-quantum physics.
There is a $\frac12 k_B T$ contribution to the energy for each independent degree of freedom:
$$
dE = \sum_n \left( \frac{1}{2} k_B T \right),
$$
where $n$ is an abstract index that labels the degrees of freedoms.
The divergence of $dE$ is the ultraviolet catastrophe that the Planck distribution remedies.
Quantum mechanics resolves the overcounting.
In asking why the vacuum energy is so small, we seek to learn how quantum gravity resolves the overcounting of the degrees of freedom
in the ultraviolet.
Similarly, in the infrared, the proper formulation of quantum theory of gravity should resolve the stability problem
(``Why doesn't the universe have a planckian size?''), once again in analogy with the resolution of the problem of atomic stability offered
by quantum mechanics. }

The question of defining a vacuum in gravity is always tied to asymptotic conditions.
We are solving eq.\ \eref{eq:einstein}, which is a differential equation, and boundary conditions are an input.
This methodology is imported to quantum gravity.
For example, in target space, string theory is formulated as an $S$-matrix theory whose long wavelength behavior is consistent with an effective field theory for gravity.
But the effective field theory in the infrared, in all cases that have been even partially understood, is particularly simple:
infinity is asymptotically flat or it is asymptotically anti-de Sitter space (AdS) or it is a plane wave limit.
The formulation of string theory as a consistent theory of quantum gravity on de Sitter space, which the present second inflationary phase of our Universe resembles, or in more general curved (time-dependent) backgrounds with (at best) approximate isometries is not at all understood.

The vacuum energy problem can also be couched in the following way.
In quantum field theory in flat space the vacuum is clearly defined as in quantum mechanics.
Vacuum energy is just the expectation value of the Hamiltonian in its ground state.
But what is the energy or Hamiltonian in a quantum theory of gravity, and what is the vacuum or lowest energy state?
Such concepts must be defined without invoking asymptopia as inputs, for clearly as local observers in spacetime, we cannot know what we are evolving towards.
How then is vacuum energy to be determined when there are no {\em a priori} fixed asymptotics and there is no Hamiltonian?

We find a clue in the equivalence principle.
In the classical theory of gravity, the nearness of a body's inertial mass to its gravitational mass is explained because an observer cannot distinguish between gravitation and acceleration.
Spacetime is locally indistinguishable from flat space (zero cosmological constant).
Globally it can be any solution at all to Einstein's equations.
It is the equivalence principle that is responsible for the dual nature of energy and the concept of a vacuum (the actual geometry of spacetime).
Because this is the root of the problem, we wish to implement the equivalence principle at the quantum mechanical level and see what light this throws on the vacuum energy problem.

In essence, we are turning the cosmological constant problem around, to argue that its natural solution
({\em i.e.}\ natural adjustment to an almost zero value)
requires a major shift in the foundations of fundamental physics.
The new fundamental postulate needed is a quantum equivalence principle which demands a consistent
gauging of the geometric structure of canonical quantum theory. %
This, we believe, is the missing key element in present formulations of consistent quantum theory of
gravity and matter. To be perfectly honest, our analysis of the cosmological constant problem within
this new framework is still largely primitive, mainly because we do not have a sufficient handle
on the new mathematical structure that is called for by the postulate of a quantum equivalence principle.

The outline of the paper is as follows.
In Section 2, we discuss how measurement differs in General Relativity and quantum mechanics.
We use this to motivate a proposal for a gauged version of quantum mechanics, whose features we review.
The canonical quantum theory we gauge is just Matrix theory.
In Section 3, we note that
the spacetime volume and the cosmological constant %
should then be regarded as conjugate quantities and fluctuate accordingly in the quantum theory.
The canonical quantum expectation value of the cosmological constant vanishes.
What is meant (and observed) by vacuum energy is the fluctuation in $\Lambda$.
We relate the smallness %
of the observed cosmological constant to the largeness of spacetime.
In Section 4, we discuss holography as it relates to our proposal.
In Section 5, we comment on the coincidence problem and indicate some future directions for this program.

\section{Quantum geometry vs.\ spacetime geometry}

We start by formulating a generalized geometric quantum theory that mimics the essential physical features of
General Relativity. In this context
we want to view the vacuum energy in the light of the peculiar non-local
properties of gravitational energy. First we discuss the nature
of observables and measurement in quantum theory and General Relativity, and then the general structure of
background independent quantum theory.\footnote{
By {\em background independence} we mean that no {\em a priori} choice of a consistent background for string propagation is made.
The usage is as in string field theory.}

\subsection{Observables and measurement in QFT and GR}

As Wigner pointed out, measurements are by nature different in quantum field theory and the General Theory of Relativity \cite{wig}.
The Special Theory of Relativity and quantum mechanics, as well as their conflation, are formulated in terms of particle trajectories or wavefunctions or fields that are functions (functionals) of positions or momenta.
Such coordinates are auxiliary constructs in the General Theory of Relativity.
Consistent with diffeomorphism invariance, we can assign almost any coordinates to label events in classical gravity and, by extension, in quantum gravity.
The coordinates are not in themselves meaningful.
Moreover, gravity is non-local, as perhaps best manifested in the concept of gravitational energy \cite{MTW}.
No local gauge invariant observables can exist \cite{einstein}.
The decoupling of scales familiar to local quantum field theories is simply not possible.
The ultraviolet (UV) physics mixes inextricably with dynamics in the infrared (IR).

Measurements in a theory of gravitation are founded upon the relational properties of spacetime events.
Timelike separation of events is measured by clocks, whereas spacelike separation is determined more indirectly.\footnote{
There are subtleties regarding the interplay between a suitably microscopic clock and a macroscopic apparatus that records the measurement \cite{swig}.}
Within a quantum theory, events cannot themselves be localized to arbitrary precision.
Only for high-energies does it even make sense to speak of a local region in spacetime where an interaction takes place.
This is a simple consequence of the energy-time uncertainty relation.

The measurements that a bulk observer makes within a quantum theory of gravity are necessarily restricted, however.
Experiments are performed in finite times and at finite scales.
The interactions accessed in the laboratory also take place in regions of low spacetime curvature.
In local quantum field theory, we operate successfully under the conceit that the light cone is rigid.
There is an approximate notion of an $S$-matrix that applies to in- and out-states with respect to the vacuum in flat space.
Computing scattering amplitudes in string theory proceeds through analytic continuation of Lorentzian spacetimes into Euclidean spaces with fixed asymptopia.
On cosmological scales, this is of course a cheat.
Light cones tilt.
The causal structure, in particular, is not static.
We do not in general know the asymptotic behavior of the metric at late times.
The only data available about spacetime are events in an observer's past light cone.
Each observer has a different past light cone consistent with the histories of all the other observers, and the future causal structure is partially inferred from these data.

A manifold is constructed out of an atlas of local coordinate charts.
A sufficiently small neighborhood about any point is flat.
To solve Einstein's equations, the vacuum energy of empty Minkowski space vanishes exactly.
Globally, the ratio of the vacuum energy density to the expectation of Planck scale physics is extremely close to zero but does not identically vanish.
We wish to regard the measured, small cosmological constant as the consequence of patching together the physics of locally flat spaces consistent with
the existence of canonical gravitational quanta.
Instead of working with the spacetime manifold, we employ a larger geometric structure whose tangent spaces are the canonical Hilbert spaces of
a consistent quantum mechanics of gravitons.
The equivalence principle we employ relies on the universality and consistency of quantum mechanics at each point.
In every small, local neighborhood of this larger structure, the notion of quantum mechanical measurement is identical.
In particular, local physics in the laboratory is decoupled from the global physics. Nevertheless, as we will emphasize in what follows, there is
a non-trivial non-decoupling of local and global physics when one discusses the quantum origin of vacuum energy. %

\subsection{Gauged quantum mechanics}
In order to implement the quantum equivalence principle, we must enlarge the framework of quantum theory.\footnote{
Generalizations of quantum mechanics have a long history \cite{generalqm, minictze, dj, geomqm, anan}.
Most recently, such generalizations were discussed, for example, in Ref.\ \cite{generalqm}.}
The standard quantum theory in these generalized frameworks is a special case of a more general theory.
We briefly summarize the proposal \cite{minictze, dj} concerning this enlargement of quantum theory via its geometric formulation.
Put simply, we gauge the unitary group of quantum theory, in the same way that the Lorentz group is gauged to the general diffeomorphism group in going from Special to General Relativity.\footnote{
The gauging of the unitary group means that in general we do not have path integrals, and thus from the point of view of the usual analogy between quantum field theory and equilibrium statistical mechanics with a Gibbsian measure, the generalized quantum theory should be of non-equilibrium type.
No real meaning assigns to the states (wavefunctions), but there is nevertheless a general dynamical statistical geometry of quantum theory.}
To explain what we mean by this, it is useful to first recall the geometrical structure of the canonical quantum theory.

It is a well known fact that standard quantum mechanics is exactly captured by the geometry of $\CP{n}$, a homogeneous, isotropic, and simply connected K\"ahler manifold with constant, holomorphic sectional curvature \cite{minictze, geomqm, anan}.
This is a maximally symmetric space.
There is a unique Riemannian metric on $\CP{n}$ inherited from the inner product on $\BC^{n+1}$ that computes the distance between two normalized states $\ket\psi$ and $\ket{\psi'} = \ket\psi + \ket{d\psi}$ to be:
\be
ds^2 = 4 \left( 1 - |\ip{\psi}{\psi'}|^2 \right) = 4\left( \ip{d\psi}{d\psi} - \ip{d\psi}{\psi}\ip{\psi}{d\psi} \right).
\label{metric}
\ee
This Cayley--Fubini--Study metric is the Fisher distance from information theory \cite{woot}.
The distance is a distance on the space of quantum events that is determined by the statistical fluctuations in the measurements performed to distinguish the first quantum state from the second.
The metric captures probabilities and the Born rule.
We may use the almost complex structure to assign coordinates $\psi_a$ and $\psi^*_a$ to $\CP{n}$.
In the $\hbar\to 0$ limit, the Fisher metric on the quantum phase space described by $\CP{n}$ reduces to a spatial metric provided that the configuration space for the quantum system under consideration is the physical space itself!
Canonical quantum theory can be viewed as a Hamiltonian system if ${\rm Re}(\psi_a)$ and ${\rm Im}(\psi_a)$ are identified with the positions and conjugate momenta $q_a$ and $p_a$, respectively.

The symplectic structure of phase space determines how wavefunctions, or more generally, fields propagate.
The Schr\"odinger equation
\be
J\hbar \frac{d}{dt} \ket{\psi(t)} = H(t) \ket{\psi(t)}
\label{sch}
\ee
is the geodesic equation with respect to the Cayley--Fubini--Study metric.
The $J$ is the almost complex structure on $\CP{n}$ defined in terms of the metric and the symplectic two-form on phase space and squares to minus the identity; it is the ``$i$'' of any quantum theory.
Using $\ket{\dot\psi}$ to denote $\frac{d}{dt}\ket\psi$, eq.\ \eref{metric} together with the Schr\"odinger equation \eref{sch} imply that
\bea
\left( \frac{ds}{dt} \right)^2 &=& 4 \left( \ip{\dot\psi}{\dot\psi} - \ip{\dot\psi}{\psi}\ip{\psi}{\dot\psi} \right) \\
&=& \frac{4}{\hbar^2} \left ( \bra{\psi} H^2 \ket{\psi} - (\bra{\psi} H \ket{\psi})^2 \right).
\label{dE2}
\eea
The expression is manifestly invariant under unitary rotation.
We recognize that the right-hand side of eq.\ \eref{dE2} is given in terms of the dispersion in energy $\Delta E$ of the Hamiltonian $H$:
\be
\Delta E = \bra{\psi} H^2 \ket{\psi} - (\bra{\psi} H \ket{\psi})^2,
\ee
and thus we see immediately that
\be
\frac{\hbar}{2}\, ds = \Delta E\, dt.
\label{time}
\ee
With a fixed Hamiltonian $H$, the evolution of an orthonormal basis for $\BC^{n+1}$ projected onto $\CP{n} \equiv U(n+1)/(U(n) \times U(1))$ is simply the geodesic equation:
\be
\frac{du^a}{ds} + \Gamma^{a}_{bc} u^b u^c = \frac{1}{2 \Delta E} {\rm Tr}(H F^a_b) u^b,
\ee
where $u^a = \frac{d}{ds} \psi^a$.
The Yang--Mills field strength $F_{ab}$ is valued in $U(n) \times U(1)$.
(A detailed exposition of this point is in, for example, Ref.\ \cite{anan}.)

The main features of quantum mechanics are embodied in the geometry of $\CP{n}$ and in the evolution equation.
The superposition principle is tied to viewing $\CP{n}$ as a collection of complex lines passing through the origin.
Entanglement arises from the embeddings of the products of two complex projective spaces within a higher dimensional one.
The geometric phase stems from the symplectic structure on $\CP{n}$.
Moreover, the expression \eref{time} relating time intervals to intervals in the projective space of the quantum theory is {\it exact}.
A similar relation between the spatial distances and geometric intervals can be established, for example
by using coherent states \cite{minictze, geomqm}, but {\it only} in the $\hbar \to 0$ limit.
This is in fact the most crucial difference between temporal and spatial geometry from the point of view of quantum geometry.
Note that in a general relativistic context, spacetime measurements can be viewed as measurements of time \cite{MTW}.
The tension between canonical quantum theory and background independent classical spacetime physics is precisely in the way the two treat measurements of time (and the corresponding canonically conjugate variable, energy).

Guided by the quantum equivalence principle, we will now generalize the structure we have developed so that the quantum geometry becomes dynamical.
The form of the Hamiltonian is fixed by the requirement that $H$ should define a canonical quantum mechanical system whose configuration space is space and whose dynamics define a consistent quantum gravity in a flat background.
We are aware of only {\em one} example of a non-perturbative quantum mechanics that satisfies this criterion: {\em Matrix theory} \cite{bfss}.
In Matrix theory, transverse space emerges as a moduli space of the supersymmetric quantum mechanics, while
time, on the other hand, appears as in any other canonical quantum theory.

By the correspondence principle, the generalized quantum geometry must locally recover the canonical quantum theory encapsulated in $\CP{n}$.
A coset of ${\rm Diff}(\BC^{n+1})$ that locally looks like $\CP{n}$ and also allows for mutually compatible metric and symplectic structures, expressed in the existence of a (generally non-integrable) almost complex structure, supplies the framework for the dynamical extension of the canonical quantum theory.
The nonlinear Grassmannian
\be
{\rm Gr}(\BC^{n+1}) = {\rm Diff}(\BC^{n+1})/{\rm Diff}(\BC^{n+1}, \BC^n\times \{0\}),
\ee
in the $n\to\infty$ limit satisfies the necessary conditions \cite{vizman}.
This space is a generalization of $\CP{n}$.
The Grassmannian is a gauged version of complex projective space, which is the geometric realization of quantum mechanics.

The utility of this formalism is that gravity embeds into quantum mechanics in the requirement that the kinematical structure must remain compatible with the generalized dynamical structure under deformation.
{\it The quantum symplectic and metric structure, and therefore the almost complex structure, are themselves fully dynamical.}
Canonical quantum mechanics applies on every local neighborhood in the space of events, and the tangent spatial transverse metric emerges from the quantum statistical metric, by ensuring that in the semiclassical $\hbar\to 0$ limit, the underlying configuration space is space itself.
The transverse space is a classical moduli space, the space of inequivalent degenerate vacua of the theory.
The longitudinal spatial coordinate corresponds to the dimensionality of the tangent Hilbert space.
Time is a measure of the geodesic distance in the space of statistical events.
Time evolution is contained in the relational properties of general information metrics.

As before, {\it locally} we have a geodesic equation that determines the time evolution of states.
This is
\be
\frac{du^a}{d\sigma} + \Gamma^{a}_{bc} u^b u^c = \frac{1}{2 M_{\rm Pl}} {\rm Tr}(H_M F^a_b) u^b,
\ee
where $H_M$ is the Matrix theory Hamiltonian,
$\Gamma^{a}_{bc}$ is the affine connection associated with the general metric $g_{ab}$, and
$F_{ab}$ is the ${\rm Diff}(\BC^{n+1},\BC^n \times \{0\})$ valued curvature two-form.
The parameter $d\sigma$ is defined via $\frac{\hbar}{2}\, d\sigma = M_{\rm Pl}\, dt$, with $M_{\rm Pl}$ the Planck energy.
(This is consistent with the energy-time uncertainty relation.)
Note that $d\sigma$ in principle contains very important conformal structure, because the effective ``charge'' in this general context is {\it not} $\frac{H_M}{\Delta E_M}$ but instead $\frac{H_M}{M_{\rm Pl}}$.
The geodesic equation follows from the conservation of the energy-momentum tensor, $\nabla_a T^{ab} = 0$, in parallel with the analogous reasoning in the context of General Relativity \cite {MTW} (Chapter 20).
Since both the metric and symplectic data are contained in $H_M$ and are $\hbar\to 0$ limits of their quantum counterparts, we have a consistent nonlinear ``bootstrap'' between the space of quantum events and the generator of the dynamics \cite{minictze, jmt}.

Physics is required to be diffeomorphism invariant in the sense of information geometry provided that the statistical metric and the symplectic structures remain compatible \cite{minictze}.
This is to say, the dynamical evolution equation of the statistical metric is the same as in the diffeomorphism invariant theory of matter and geometry:
\begin{equation}
\label{BIQM1}
R_{ab} - \frac{1}{2} g_{ab} R  - \lambda g_{ab} = T_{ab} (F_{ab}, H_M).
\end{equation}
This is the Einstein--Yang--Mills equation on the space of quantum events.
(The constant term $\lambda = (n+1)/\hbar$ for $\CP{n}$, and in this case $M_{\rm Pl}\to \infty$.)
Moreover, we must demand for compatibility
\begin{equation}
\label{BIQM2}
\nabla_a F^{ab} = \frac{1}{M_{\rm Pl}} H_M u^b.
\end{equation}
The two equations imply, via the Bianchi identity, a conserved energy-momentum tensor, which together with the conserved ``current''
$j^a \equiv \frac{1}{2M_{\rm Pl}} H_M u^a$, $\nabla_a j^a =0$,
results in the generalized ``non-linear'' geodesic Schr\"{o}dinger equation.
As in General Relativity, it is crucial to understand both local and global features of solutions to the dynamical equations.\footnote{
Based on general symmetry considerations, the gauged Matrix theory is compatible with other non-perturbative formulations of string theory in curved backgrounds, most notably in the AdS/CFT correspondence \cite{ads}.
The dual CFT, in the case of $16$ supercharges, is reducible (upon dimensional reduction to $0+1$ dimensions) to Matrix theory.
The Matrix theory must capture the physics
of local flat regions of the AdS space, which according to the principle of equivalence, are physically independent from the AdS asymptotics.
Thus there is no conflict between gauged Matrix theory and AdS/CFT from the global point of view, but the gauged Matrix theory offers a more general formulation of string theory, because it can in principle be applied to other backgrounds, such as time-dependent cosmological situations, in which case it is not necessarily physically meaningful to adhere to the notions of fixed asymptopia, or global holographic screens, or $S$-matrix observables, or dual CFTs.}

When the metric and symplectic form on phase space become fully dynamical, we conclude that only individual quantum events make any sense observationally.
Moreover, the requirement of diffeomorphism invariance places stringent constraints on the quantum geometry.
We must have a strictly ({\em i.e.}\ non-integrable) almost complex structure on the generalized space of quantum events.
This extended framework readily implies that the wavefunctions labeling the event space, while still unobservable, are in fact irrelevant.
They are as meaningless as coordinates in General Relativity.
The physics does not rely on such choices.
At the basic level, there are only dynamical correlations of quantum events.
Observables are furnished by diffeomorphism invariant quantities in the quantum configuration space.

One important element of this approach to quantum gravity is the existence of a correspondence limit between the dynamical quantum theory and the classical Einstein theory of gravity coupled to matter.
At long wavelengths, once we map the configuration space for the states to spacetime, we have General Relativity.
Turning off dynamics in the quantum configuration space recovers the canonical quantum mechanics.
Gauging the quantum mechanics generically breaks supersymmetry.
We do not have globally defined supercharges in spacetime in the correspondence limit.
This explains why there is a cosmological constant.

\section{Why the cosmological constant is small}

General quantum theory implies that
the vacuum energy is a dynamical observable, and not a number (or coupling to
be evaluated at some energy scale). Thus it is legitimate to talk about fluctuations of
vacuum energy. The crucial point of generalized quantum theory is that in its context it is
possible to make the canonical expectation value for the vacuum energy vanish in local regions of spacetime, as well as globally.
This crucially follows from the quantum equivalence principle, in complete analogy with the general relativistic treatment of gravitational
energy based on the classical principle of equivalence (see \cite{MTW}, Chapter 19).
In this section we use a semiclassical limit to relate fluctuations of vacuum energy to its
canonically conjugate quantity --- the effective spacetime volume.
Matrix theory provides a UV/IR
relation
to correlate the ultraviolet and infrared scales via the readings of dynamical quantum clocks as
implied by general quantum theory.
Following the quantum equivalence principle, we
end up with a locally flat spacetime (whose stability will be
ensured by the supersymmetric quantum mechanics we are gauging), independent of global structure.
For a closed universe, one has a vanishing canonical expectation value
for the vacuum energy, yet naturally small fluctuations of the same.

\subsection{$\Lambda$ vs.\ V}
At each point in the spacetime manifold, the space is locally flat.
Locally, the vacuum energy is fixed by the quantum theory in the tangent space, which is Matrix theory.
The structure of the light cones is codified in the matrices \cite{bfss}.
The unbroken supersymmetry in Matrix theory ensures that the vacuum energy vanishes identically within each given local neighborhood.
Quasi-locally, the vacuum weighs nothing; its expectation value is zero.
This must be so for Minkowski space to be a solution to the Einstein equations.
The obstruction to extending this into a global statement for the manifold as a whole arises from patching together the physics of the tangent spaces at different points in spacetime.
The observed value of the cosmological constant has a natural interpretation as a fluctuation about the zero mean.
The dynamical, cosmological breaking of supersymmetry gives vacuum a weight.

In string theory, at least semiclassically, the spacetime is ${\cal M}_4 \times {\cal K}_6$, where ${\cal M}_4$ is the observed, macroscopic spacetime, and ${\cal K}_6$ is a compact space, such as Calabi--Yau threefold, whose details will not be important for our purposes.
{\em The smallness of the observed cosmological constant $\Lambda$ is a statement about the largeness of the manifold ${\cal M}_4$.}

In semiclassical gravity, the cosmological constant arises in the Einstein--Hilbert action as a prefactor for the volume of the
four-dimensional\footnote{In Section 4, we will see that there is something special about {\it four} spacetime dimensions!} spacetime ${\cal M}_4$:
\be
V = \int d^4x\ \sqrt{-g}.
\ee
As it is the product $\Lambda \cdot V$ that appears, $\Lambda$ and $V$ should be regarded as canonically conjugate quantities.
In a quantum theory, we expect that the fluctuation in one observable is related to fluctuations in its conjugate.
Thus (see also \cite{by, pad, vol, euro}),
\begin{equation}
\Delta\Lambda\ \Delta V \sim \hbar.
\label{eq:fluc}
\end{equation}
This is an energy-time uncertainty relation in the spacetime (the string theory target space).
The preferred value of the cosmological constant is identically zero.
The existence of a measured vacuum energy is the consequence of quantum fluctuations about the zero value.
Fluctuations in $\Lambda$ are inversely related to fluctuations in $V$.

Fluctuations in the volume of spacetime are fixed by statistical fluctuations in the number of degrees of freedom of the gauged quantum mechanics.
In Matrix theory, the eigenvalues of the matrices denote the positions of D$0$-branes which give rise to coherent states in gravity \cite{bfss}.
Off-diagonal terms in Matrix theory break the permutation symmetry and render the D$0$-branes distinguishable.
Therefore, to enumerate the degrees of freedom, we employ the statistics of distinguishable particles.
The fluctuation is given by a Poisson distribution, which is typical for coherent states.

The fluctuation of relevance for us is in the number of Planck sized cells that fill up the configuration space (the space in which quantum events transpire).
That is to say,
\be
\CN_{\rm cells} \sim \frac{V}{\ell_{\rm Pl}^D} \Longrightarrow
\Delta \CN_{\rm cells} \sim \sqrt{\CN_{\rm cells}} \Longrightarrow
\Delta V \sim \sqrt{V}\ \ell_{\rm Pl}^{D/2}.
\ee
In $D$ spacetime dimensions, the Newton constant $G_D$ and the Planck constant $\hbar$ go as
\bea
G_D \sim \ell_{\rm Pl}^{D-3}\ M_{\rm Pl}^{-1}, &&
\hbar \sim \ell_{\rm Pl}\ M_{\rm Pl}.
\eea
Thus, eq.\ \eref{eq:fluc} informs us that
\be
\Delta\Lambda \sim \frac{\hbar}{\Delta V}
\sim \frac{\hbar^{(D-4)/2(D-2)}}{\sqrt{V}\ G_D^{D/2(D-2)}}.
\label{eq:deltaL}
\ee
In particular for $D=4$, we have
\be
\Delta\Lambda \sim \frac{1}{\sqrt{V}\ G_4}.
\ee
It is this $\Delta\Lambda$ that is the observed vacuum energy or cosmological constant.
Because the volume of spacetime is large, the cosmological constant is small.

Finally, we need to address the question of the sign of the fluctuations of the vacuum energy.\footnote{We thank Don Marolf for a crucial
comment regarding this issue.}
{\em A priori}, it seems that there is no restriction of the sign of the fluctuation, in which case one might face serious observational problems
(\cite{by}, last reference). In our context, the positive sign of vacuum energy fluctuations might be favored in the probability distribution
for $\Lambda$
that is in principle computable in the ``non-linear'' and ``non-equilibrium'' context of generalized quantum theory.
Based on the general considerations, we expect a probability distribution that is a corrected Gaussian (as in typical ``non-equilibrium''
considerations), consistent with the
fluctuation estimate presented above. A computation of this probability would be also crucial for the understanding of the coincidence
problem, as we briefly allude to at the end of this paper.

\subsection{UV vs.\ IR}
How small is the cosmological constant?
To answer this question it is important to remember that the cosmological constant connects the ultraviolet (vacuum degrees of freedom) and the infrared (large-scale dynamics).
The spacetime uncertainty relation \cite{ur} in string theory accounts for the mixing of scales.
In perturbative string theory, modular invariance on the worldsheet translates in target space to the spacetime uncertainty relation:
\be
\Delta T \, \Delta X_{\rm tr} \sim \ell_s^2 \sim \alpha'.
\ee
(Here, $T$ is a timelike direction, and $X_{\rm tr}$ is a spacelike direction transverse to the lightcone.)
Scales on configuration space --- and thus in the semiclassical limit, on the string target space --- are determined by $\Delta T$, which provides a clock in General Relativity through which relative distances between events are measured.
The spacetime uncertainty relation is a statement about conformal invariance and consistency of the worldsheet theory.
This is a statement about the internal space and is logically distinct from the uncertainty relation between $\Delta V$ and $\Delta\Lambda$.

The spacetime uncertainty relation from perturbative string theory lifts to M-theory.
To be precise, we consider the strong coupling limit of type IIA string theory \cite{ew}.
Non-perturbatively, the eleven-dimensional Planck length $\ell_{\rm Pl} = \hbar\ M_{\rm Pl}^{-1}$ and the size of the M-theory circle that extends at large string coupling set the string scale $\alpha' = \ell_{\rm Pl}^3/r$.
The radius $r = g_s^{1/3} \ell_s$ of the circle determines the maximal uncertainty in $X_{\rm long}$, the longitudinal direction in the Matrix theory limit, which implies that
\be
\Delta T \, \Delta X_{\rm tr} \, \Delta X_{\rm long} \sim \ell_{\rm Pl}^3,
\ee
a cubic relation consistent with the existence of membranous structures in M-theory \cite{ur}.
The Planck energy and the geodesic distance on the space of quantum events are related by
\be
\hbar\, \Delta s \sim M_{\rm Pl}\, \Delta T.
\label{eq:deltas}
\ee
Although $\hbar$ and $M_{\rm Pl}$ are given, dimensionful numbers, the time dependence is not {\em a priori} fixed.
Generalizing what we have already said, we have
$
\hbar\,ds = \Delta E_g\,dt,
$
where the dispersion in energy in the presence of a
gravitational coupling for a given matter Hamiltonian is determined by
$
\Delta E = \bra{\psi} H_g^2 \ket{\psi} - (\bra{\psi} H_g \ket{\psi})^2
$
where $H_g$ is the matter Hamiltonian in a given gravitational background.
For example, the known redshift formula in the constant gravitation field comes out immediately:
physically, we have a quantum clock in a gravitational field, which is how the redshift is
determined experimentally --- once again, measurements in spacetime are always measurements of time,
in General Relativity \cite{MTW}.
Note that here we have a complete analogy between the known fact that gravitational redshift demands
curved spacetime \cite{MTW}, and that quantum gravitationally shifted vacuum energy demands curved
quantum geometry.

In eq.\ \eref{eq:deltas},
$\Delta s$ is dynamical and can grow or shrink, depending upon the background.
The geodesic distance in the gauged quantum mechanics contains a dynamical conformal factor, to account for the
dynamical causal structure (arising from the compatibility of the generalized dynamical metrical and symplectic
structures of quantum theory).
The size of spacetime is obtained in the semiclassical limit by the measure on the space of quantum events.

\subsection{Gravitational see-saw}
Unlike the usual formulation of quantum mechanics, this discussion is not couched as a dispersion relation for the Hamiltonian.
When a photon falls in a uniform gravitational field, the Hamiltonian acquires a gravitational potential.
This feeds into the dispersion relation, which alters the measurement of the clock.
The conformal factor we have introduced is a redshift parameter that serves the identical purpose.
This bears an analogy to the Randall--Sundrum constructions in which an exponential warp factor explains the hierarchy between electroweak and fundamental scales \cite{rs1}.
The redshift factor similarly explains the hierarchy between the cosmological constant and the Planck energy.

We estimate the line element on the space of probabilities to scale as
\be
ds \sim e^{-S_{\rm eff}},
\ee
where $S_{\rm eff}$ denotes the low-energy (Euclidean) effective action for the matter degrees of freedom propagating in an emergent (fixed) spacetime background.
This is the usual estimate based on the canonical quantum theory, as formulated in terms of path integrals, in a semi-classical limit.

The spacetime uncertainty relation becomes
\begin{equation}
\Delta X_{\rm tr} \, \Delta X_{\rm long} \sim e^{S_{\rm eff}}\, \ell_{\rm Pl}^2.
\end{equation}
The product of the ultraviolet cutoff (the maximal uncertainty in the transverse coordinate) and the infrared cutoff (the maximal uncertainty in the longitudinal coordinate) is thus exponentially suppressed compared to the Planck scale.
There is a {\em gravitational see-saw} \cite{jmt}.
Within an emergent spacetime background, at low energies, we locally have a supergravity theory, but there is in general no globally defined supersymmetry.
Moreover, a timelike Killing direction is only a local notion.

All the physics is in the effective action $S_{\rm eff}$ obtained from dimensional reduction of the matrix formulation of M-theory on the appropriate background.
Our discussion regarding $S_{\rm eff}$ is qualitative rather than explicit for two reasons.
Firstly, a non-perturbative technology for treating Matrix theory in an arbitrary background is presently lacking.
Secondly, our mathematical knowledge about the geometric structure of the infinite dimensional Grassmannian ${\rm Gr}(\BC^n)$ in the $n\to\infty$ limit, which defines the generalized quantum mechanics over the full spacetime manifold, is incomplete.
An improved understanding of these structures is necessary to lend rigor to the scaling arguments that we make.

As we have already pointed out, gauging the geometric structure of canonical quantum mechanics generically breaks supersymmetry.
 From $\Delta X_{\rm tr} \, \Delta X_{\rm long} \sim e^{S_{\rm eff}}\, \ell_{\rm Pl}^2$, it follows that the effective scale of (in general, cosmological) supersymmetry breaking $m_{\rm susy}$ provides an estimate for the vacuum energy in four dimensions as implied by the gravitational see-saw:
\be
\omega^4 \equiv \Delta\Lambda \sim \left( \frac{m_{\rm susy}^2}{M_{\rm Pl}} \right)^4.
\ee
This is how much the vacuum weighs \cite{jlm}.
The cosmological breaking of supersymmetry argues the intimate connection between the volume of spacetime and the observed vacuum energy \cite{banks}.
In the $\omega \to 0$ limit, we recover supersymmetry in spacetime and diffeomorphism invariance in the space of quantum events.
The numerical smallness of the vacuum energy density is thus consistent with the principle of naturalness in that the vanishing of a dimensionful quantity restores a dynamically broken symmetry. Note that in this approach the non-perturbative stability of local Minkowski regions (as needed by quantum equivalence principle) is due to the fact that we are gauging
a supersymmetric quantum theory, which is also responsible for the local vanishing of vacuum energy.

\section{Comment on holography}
It is well known that holography provides heuristic support for a cosmological constant far smaller than the exaggerated expectations of effective field theory.
According to holography, the degrees of freedom of gravity in $D$ spacetime dimensions are captured by equivalent non-gravitational physics in $D-1$ dimensions \cite{holo}.
The relation between holography and the cosmological constant was explored in Refs.\ \cite{tb, ckn, hm}.

To be precise, we briefly recall Thomas's holographic analysis of the cosmological constant \cite{thomas}.
Suppose there are $D$ spacetime dimensions, each with a characteristic scale $R$.
The holographic bound demands that the entropy (the number of degrees of freedom $\CN_{\rm dof}$) scales as the area \cite{bh}.
This is a $(D-2)$-dimensional surface, so
\begin{equation}
\CN_{\rm dof} \leq \frac{R^{D-2}}{4 G_D \hbar}.
\end{equation}
 From the uncertainty principle, the energy of each independent degree of freedom scales as
\begin{equation}
E \sim \frac{\hbar}{R}.
\end{equation}
All the degrees of freedom contribute equally to the vacuum energy density:
\begin{equation}
\Lambda \sim \CN_{\rm dof} \frac{E}{R^{D-1}}
\sim \frac{1}{R^2 G_D}.
\label{eq:scott}
\end{equation}
The dependence on $\hbar$ has cancelled in the last expression, so this vacuum energy density should survive the semiclassical ($\hbar\to 0$) limit.
Here, the cosmological constant $\Lambda$ is a prescribed, fixed number.
It is determined by the size $R$ of the regularized ``box.'' Note that this is somewhat na\"{\i}ve, because the characteristic sizes
of spatial and temporal directions do not have to be the same, as discussed in the previous section.

We notice that the fluctuation $\Delta\Lambda$ that we obtain in eq.\ \eref{eq:deltaL} matches this scaling, but only when $D=4$.
A Poisson fluctuation in the holographic degrees of freedom $\CN_{\rm dof}$ will not recover the holographic scaling of the cosmological constant.
The fluctuation we have considered is in the volume of the quantum mechanical configuration space rather than in a codimension one structure.
The reason that this is consistent is that we have applied the principle of equivalence at each point in spacetime.
In the scheme that we have proposed, holography enters in the choice of the quantum theory compatible with having Minkowski space as a local solution, namely through Matrix theory.
Holography is ``local,'' in the sense of the equivalence principle.
There does seem to be some tension, however, with global holography, which might be a useful concept only for certain
physics questions.

\section{Lessons of $\Lambda$}
Because the future causal structure of spacetime is unknown, a global $S$-matrix description of our Universe is unavailable.
 From the viewpoint of the generalized quantum theory, a wavefunctional approach \`a la Wheeler--DeWitt \cite{wdw} seems poorly formulated.
In the general quantum theory we can have dynamical statistical correlations between the past and today.
The observables of quantum gravity are these dynamical correlations in the configuration space of the quantum theory.
A functional approach to quantum gravity is to consider a canonical theory of quantum mechanics (Matrix theory) at every point in spacetime, where spacetime is here regarded as a semiclassical geometry that arises from identifying the configuration space with the physical space in the $\hbar\to 0$ limit.
This is an application of the correspondence principle that ensures that at long wavelengths we recover General Relativity.

In gauging quantum mechanics to lift the ten-dimensional (or eleven-dimensional) vacuum, we obtain a vacuum energy that corresponds to the cosmological scale of supersymmetry breaking.
The second cosmological constant problem nevertheless persists:
why is $\Omega_\Lambda \approx \Omega_{\rm matter}$ {\em today}?
This could be an accident of living in the present epoch \cite{anthropics}.
We would, in view of the main point of this paper, prefer to view the cosmic concordance problem through a different dynamical lens, one might want to term ``the universe as an attractor.''

The classic reference \cite{msw} considers linear perturbation theory in a Friedmann--Robertson--Walker (FRW) background for a certain density of matter and a certain vacuum energy and and then deduces an {\em a priori} probability for the vacuum energy so that
\begin{enumerate}
\item gravitational bound states appear at large scales;
\item the fundamental constants are held fixed; and
\item the probability distribution is independent of a bare vacuum energy, which permits the use of Bayesian statistics.
\end{enumerate}
It is in holding the constants fixed that considerations based on observer bias, {\em i.e.}\ anthropic selection, takes place.
We may be able to apply a similar reasoning with generalized quantum mechanics, but without resorting to anthropic selection.
In the framework of gauged quantum mechanics, non-Gibbsian quantum probability distributions are dynamically possible, for example as perturbations of the usual path integral, around the Fisher metric.
Anthropic reasoning is evaded because we have an $S_{\rm eff}$ that can in principle be obtained directly and exactly from Matrix theory.
Thus a dynamical resolution of the coincidence problem might be possible.
We need to explore ``non-linear'' and ``non-equilibrium'' features of generalized quantum theory in order to examine this possibility in detail.
This is work in progress.

\section*{Acknowledgments}
We thank many colleagues for their comments. Most recently we have enjoyed discussions with Damien Easson, Laurent Freidel,
James Gray, Mike Kavic, Rob Leigh, Don Marolf, Tatsu Takeuchi, Chia-Hsiung Tze, and various participants
of the William and Mary conference ``Beyond the Standard Model''.
Special thanks to Jan de Boer and Don Marolf for many thoughtful comments on a preliminary version of this
paper.
VJ is supported by PPARC.
DM is supported in part by the U.S.\ Department of Energy under contract DE-FG05-92ER40677.
Finally, we thank the Kavli Institute for Theoretical Physics and Perimeter Institute for hospitality.


\begin{thebibliography}{99}
{\small

\bibitem{sn}
A.~G.~Riess, {\it et al.}\ [Supernova Search Team Collaboration],
``Observational evidence from supernovae for an accelerating universe and a cosmological constant,''
Astron.\ J.\  {\bf 116}, 1009 (1998)
[astro-ph/9805201];
S.~Perlmutter, {\it et al.}\ [Supernova Cosmology Project Collaboration],
``Measurements of Omega and Lambda from 42 high-redshift supernovae,''
Astrophys.\ J.\  {\bf 517}, 565 (1999)
[astro-ph/9812133].

\bibitem{wmap3}
D.~N.~Spergel, {\it et al.}\ [WMAP Collaboration],
``Wilkinson Microwave Anisotropy Probe (WMAP) Three year results: Implications for cosmology,''
astro-ph/0603449.
%
%
%
%
%
%
%
%

\bibitem{weinb}
For reviews, see
S.~Weinberg,
``The cosmological constant problem,''
Rev.\ Mod.\ Phys.\  {\bf 61}, 1 (1989);
%
``The cosmological constant problems,''
astro-ph/0005265.

\bibitem{wig}
E.~P.~Wigner,
``Relativistic invariance and quantum phenomena,''
Rev.\ Mod.\ Phys.\  {\bf 29}, 255 (1957).

\bibitem{MTW}
C.~W.~Misner, K.~S.~Thorne, and J.~A.~Wheeler,
{\it Gravitation}, Chs.\ 7, 16, 19, 20; New York: W.~H.~Freeman (1973).

\bibitem{einstein}
Interestingly, this was well known from the beginning:
A.~Einstein, {\em Albert Einstein and Michele Besso Correspondence}, letter of 3 January 1916, ed.\ P.~Speziali, Paris: Hermann (1972);
{\em Relativity: the Special and General Theory}, New York: Dover (2001).
For a recent discussion of these issues, see
S.~B.~Giddings, D.~Marolf, and J.~B.~Hartle,
``Observables in effective gravity,''
hep-th/0512200.


\bibitem{swig}
H.~Salecker and E.~P.~Wigner,
``Quantum limitations of the measurement of space-time distances,''
Phys.\ Rev. {\bf 109}, 571 (1958).

\bibitem{generalqm}
See, for example, S.~L.~Adler, {\it Quantum theory as an emergent phenomenon: The statistical mechanics of matrix models as the precursor of quantum field theory},
Cambridge, University Press (2004); J.~B.~Hartle,
``Generalizing quantum mechanics for quantum spacetime,''
  gr-qc/0602013;
A.~J.~Leggett, ``Testing the limits of Quantum mechanics: Motivation, State-of-Play, Prospects,'' J.\ Phys.\ Cond.\ Matt.\ {\bf 14},
R415 (2002); R.~Penrose,
``Wavefunction collapse as a real gravitational effect,''
in {\it Mathematical Physics 2000}, A.~Fokas, {\it et al.}, eds., London: Imperial College Press, 266;
G.~'t~Hooft, ``Determinism beneath quantum mechanics,'' quant-phys/0212095, and references therein.

\bibitem{minictze}
D.~Minic and C.~H.~Tze,
``Background independent quantum mechanics and gravity,''
Phys.\ Rev.\ D {\bf 68}, 061501 (2003)
[hep-th/0305193];
%
``Quantum mechanics as a theory of principles, and beyond,''
Phys.\ Lett.\ B {\bf 581}, 111 (2004)
[hep-th/0309239];
%
``What is quantum theory of gravity?,''
hep-th/0401028;
V.~Jejjala, M.~Kavic, and D.~Minic, ``Time and M-theory'', a review, in preparation.

\bibitem{dj}
The work in Ref.\ \cite{minictze} is a continuation of
D.~Minic and C.~H.~Tze,
``Nambu quantum mechanics: A nonlinear generalization of geometric quantum mechanics,'' %
Phys.\ Lett.\ B {\bf 536}, 305 (2002)
[hep-th/0202173];
H.~Awata, M.~Li, D.~Minic and T.~Yoneya,
``On the quantization of Nambu brackets,''
JHEP {\bf 0102}, 013 (2001)
[hep-th/9906248];
D.~Minic,
``Towards covariant matrix theory,''
hep-th/0009131;
%
``M-theory and deformation quantization,''
hep-th/9909022.

\bibitem{geomqm}
B.~Mielnik,
``Generalized quantum mechanics,''
Comm.\ Math.\ Phys.\ {\bf 37}, 221 (1974);
T.~W.~Kibble,
``Geometrization of quantum mechanics,''
Comm.\ Math.\ Phys.\  {\bf 65}, 189 (1979);
J.~P.~Provost and G.~Vallee,
``Riemannian structure on manifolds of quantum states,''
Comm.\ Math.\ Phys.\ {\bf 76}, 289 (1980);
A.~Heslot,
``Quantum mechanics as a classical theory,''
Phys.\ Rev.\ D {\bf 31}, 1341 (1985);
A.~M.~Bloch,
``An infinite-dimensional classical integrable system and the Heisenberg and Schr\"odinger representations,''
Phys.\ Lett.\ A {\bf 116}, 353 (1986);
J.~Anandan and Y.~Aharonov,
``Geometry of quantum evolution,''
Phys.\ Rev.\ Lett.\  {\bf 65}, 1697 (1990);
R.~Cirelli, A.~Mania, and I.~Pizzocchero,
``Quantum mechanics as an infinite-dimensional Hamiltonian system with uncertainty structure, I, II,''
J.\ Math.\ Phys.\ {\bf 31}, 2891 (1990);
S.~Weinberg,
``Testing quantum mechanics,''
Annals Phys.\  {\bf 194}, 336 (1989);
J.~Polchinski,
``Weinberg's nonlinear quantum mechanics and the EPR paradox,''
Phys.\ Rev.\ Lett.\  {\bf 66}, 397 (1991);
G.~W.~Gibbons,
``Typical states and density matrices,''
J.\ Geom.\ Phys.\  {\bf 8}, 147 (1992);
G.~W.~Gibbons and H.~J.~Pohle,
``Complex numbers, quantum mechanics, and the beginning of time,''
Nucl.\ Phys.\ B {\bf 410}, 117 (1993)
[gr-qc/9302002];
A.~Ashtekar and T.~A.~Schilling,
``Geometrical formulation of quantum mechanics,''
gr-qc/9706069;
D.~C.~Brody and L.~P.~Hughston,
``Geometric quantum mechanics,''
J.\ Geom.\ Phys.\  {\bf 38}, 19 (2001)
[quant-ph/9906086];
N.~P.~Landsmann,
``Classical behaviour in quantum mechanics: a transition probability approach,''
Int.\ Jour.\ Mod.\ Phys.\ {\bf B 10}, 1545 (1996)
[quant-ph/9511001];
``Poisson spaces with a transition probability,''
Rev.\ Math.\ Phys.\ {\bf 9}, 29 (1997)
[quant-ph/9603005];
{\it Mathematical topics between classical and quantum mechanics},
Springer, New York (1998);
and references therein.

\bibitem{anan}
J.~Anandan,
``A geometric approach to quantum mechanics,''
Found.\ Phys.\ {\bf 21}, 1265 (1991).

\bibitem{woot}
W.~K.~Wootters,
``Statistical distance and Hilbert space,''
Phys.\ Rev.\ D {\bf 23}, 357 (1981).


\bibitem{bfss}
T.~Banks, W.~Fischler, S.~H.~Shenker, and L.~Susskind,
``M theory as a matrix model: A conjecture,''
Phys.\ Rev.\ D {\bf 55}, 5112 (1997)
[hep-th/9610043].


\bibitem{vizman}
S.~Haller and C.~Vizman,
``Non-linear Grassmannian as coadjoint orbits,''
math.DG/0305089.


\bibitem{jmt}
V.~Jejjala, D.~Minic, and C.~H.~Tze,
``Toward a background independent quantum theory of gravity,''
Int.\ J.\ Mod.\ Phys.\ D {\bf 13}, 2307 (2004)
[gr-qc/0406037].


\bibitem{ads}
J.~M.~Maldacena,
``The large N limit of superconformal field theories and supergravity,''
Adv.\ Theor.\ Math.\ Phys.\  {\bf 2}, 231 (1998)
[Int.\ J.\ Theor.\ Phys.\  {\bf 38}, 1113 (1999)]
[hep-th/9711200];
S.~S.~Gubser, I.~R.~Klebanov, and A.~M.~Polyakov,
``Gauge theory correlators from non-critical string theory,''
Phys.\ Lett.\ B {\bf 428}, 105 (1998)
[hep-th/9802109];
E.~Witten,
``Anti-de Sitter space and holography,''
Adv.\ Theor.\ Math.\ Phys.\  {\bf 2}, 253 (1998)
[hep-th/9802150].

\bibitem{by}
Sorkin has pioneered a similar argument in a completely different context in his work on causal sets:
R.~D.~Sorkin,
``First steps with causal sets,''
in {\it General Relativity and Gravitational Physics},
R.~Cianci, {\it et al}, eds.,
Singapore: World Scientific (1991);
``Spacetime and causal sets,''
in {\it Relativity and Gravitation:  Classical and Quantum},
J.~C.~D'Olivo, {\it et al}, eds.,
Singapore: World Scientific (1991);
`Discrete gravity,''
lectures to
{\it First Workshop on Mathematical Physics and Gravitation},
held in Oaxtepec, Mexico (1995);
``A review of the causal set approach to quantum gravity''
and ``A growth dynamics for causal sets,''
lectures to
{\it New Directions in Simplicial Quantum Gravity},
held in Santa Fe (1997);
``Forks in the road, on the way to quantum gravity,''
Int.\ J.\ Theor.\ Phys.\  {\bf 36}, 2759 (1997)
[gr-qc/9706002];
``Causal sets: Discrete gravity,''
gr-qc/0309009;
``Big extra dimensions make Lambda too small,''
Braz.\ J.\ Phys.\  {\bf 35}, 280 (2005)
[gr-qc/0503057].
In particular, consult
M.~Ahmed, S.~Dodelson, P.~B.~Greene, and R.~Sorkin,
``Everpresent Lambda,''
Phys.\ Rev.\ D {\bf 69}, 103523 (2004),
[astro-ph/0209274], and references therein.

\bibitem{pad}
T.~Padmanabhan,
``Vacuum fluctuations of energy density can lead to the observed cosmological constant,''
Class.\ Quant.\ Grav.\  {\bf 22}, L107 (2005)
[hep-th/0406060];
``Dark Energy: Mystery of the Millennium,''
astro-ph/0603114.

\bibitem{vol}
For related discussion, see also
G.~E.~Volovik,
``Vacuum energy: Myths and reality,''
gr-qc/0604062;
``Thermodynamic fluctuations of Lambda and estimation of the size of universe,''
JETP Lett.\  {\bf 80}, 465 (2004)
[gr-qc/0406005];
``Cosmological constant and vacuum energy,''
Annalen Phys.\  {\bf 14}, 165 (2005)
[gr-qc/0405012].

\bibitem{euro}
Interestingly, the following article makes a similar point about hadronic vacuum energy:
A.~Aurilia and E.~Spallucci,
``Quantum fluctuations of a 'constant' gauge field,''
Phys.\ Rev.\ D {\bf 69}, 105004 (2004)
[hep-th/0402096].

\bibitem{ur}
M.~Li and T.~Yoneya,
``D-particle dynamics and the space-time uncertainty relation,''
Phys.\ Rev.\ Lett.\  {\bf 78}, 1219 (1997)
[hep-th/9611072];
T.~Yoneya,
``Schild action and space-time uncertainty principle in string theory,''
Prog.\ Theor.\ Phys.\  {\bf 97}, 949 (1997)
[hep-th/9703078];
M.~Li and T.~Yoneya,
``Short-distance space-time structure and black holes in string theory: A short review of the present status,''
hep-th/9806240;
D.~Minic,
``On the space-time uncertainty principle and holography,''
Phys.\ Lett.\ B {\bf 442}, 102 (1998)
[hep-th/9808035].

\bibitem{ew}
E.~Witten,
``String theory dynamics in various dimensions,''
Nucl.\ Phys.\ B {\bf 443}, 85 (1995)
[hep-th/9503124].

%
%
%
%
%

\bibitem{rs1}
L.~Randall and R.~Sundrum,
``A large mass hierarchy from a small extra dimension,''
Phys.\ Rev.\ Lett.\  {\bf 83}, 3370 (1999)
[hep-ph/9905221].

\bibitem{jlm}
This behavior also emerges in the deconstruction of four-dimensional gravity:
V.~Jejjala, R.~G.~Leigh, and D.~Minic,
``The cosmological constant and the deconstruction of gravity,''
Phys.\ Lett.\ B {\bf 556}, 71 (2003)
[hep-th/0212057];
%
``Deconstruction and holography,''
JCAP {\bf 0306}, 002 (2003)
[hep-th/0302230];
%
``Deconstructing the cosmological constant,''
Gen.\ Rel.\ Grav.\  {\bf 35}, 2089 (2003)
[gr-qc/0305072]. These papers were based on the proposal
of E.~Witten,
  ``Is supersymmetry really broken?,''
  Int.\ J.\ Mod.\ Phys.\ A {\bf 10}, 1247 (1995)
  [hep-th/9409111];
  ``Strong coupling and the cosmological constant,''
  Mod.\ Phys.\ Lett.\ A {\bf 10}, 2153 (1995)
  [hep-th/9506101].

\bibitem{banks}
T.~Banks,
``Cosmological breaking of supersymmetry or little Lambda goes back to the future, II,''
hep-th/0007146.

\bibitem{holo}
G.~'t Hooft,
``Dimensional reduction in quantum gravity,''
gr-qc/9310026;
L.~Susskind,
``The world as a hologram,''
J.\ Math.\ Phys.\  {\bf 36}, 6377 (1995)
[hep-th/9409089].

\bibitem{tb}
T.~Banks,
``SUSY breaking, cosmology, vacuum selection and the cosmological constant in string theory,''
hep-th/9601151.

\bibitem{ckn}
A.~G.~Cohen, D.~B.~Kaplan, and A.~E.~Nelson,
``Effective field theory, black holes, and the cosmological constant,''
Phys.\ Rev.\ Lett.\  {\bf 82}, 4971 (1999)
[hep-th/9803132].

\bibitem{hm}
P.~Horava and D.~Minic,
``Probable values of the cosmological constant in a holographic theory,''
Phys.\ Rev.\ Lett.\  {\bf 85}, 1610 (2000)
[hep-th/0001145].

\bibitem{thomas}
S.~D.~Thomas,
``Holography stabilizes the vacuum energy,''
Phys.\ Rev.\ Lett.\  {\bf 89}, 081301 (2002);
%
``Holographic vacuum energy,''
hep-th/0010145.

\bibitem{bh}
J.~D.~Bekenstein,
``Black holes and entropy,''
Phys.\ Rev.\ D {\bf 7}, 2333 (1973);
J.~M.~Bardeen, B.~Carter, and S.~W.~Hawking,
``The four laws of black hole mechanics,''
Commun.\ Math.\ Phys.\  {\bf 31}, 161 (1973);
S.~W.~Hawking,
``Particle creation by black holes,''
Commun.\ Math.\ Phys.\  {\bf 43}, 199 (1975)
[Erratum-ibid.\  {\bf 46}, 206 (1976)].

\bibitem{wdw}
B.~S.~DeWitt,
``Quantum theory of gravity. 1. The canonical theory,''
Phys.\ Rev.\  {\bf 160}, 1113 (1967);
J.~A.~Wheeler,
``Superspace and the nature of quantum geometrodynamics,''
in {\em Battelle Rencontres: 1967 Lectures in Mathematics and Physics},
C.~DeWitt and J.~A.~Wheeler, eds.,
New York: W.~A.~Benjamin (1968).

\bibitem{anthropics}
For recent discussion of the anthropic principle, see
R.~Bousso and J.~Polchinski,
``Quantization of four-form fluxes and dynamical neutralization of the cosmological constant,''
JHEP {\bf 0006}, 006 (2000)
[hep-th/0004134];
L.~Susskind,
``The anthropic landscape of string theory,''
hep-th/0302219;
and references therein. A very nice recent review is
 J.~Polchinski,
  ``The cosmological constant and the string landscape,''
  hep-th/0603249.
The notion of tuning the cosmological constant has a long history.
See, for example,
A.~D.~Linde,
``Chaotic inflation,''
Phys.\ Lett.\ B {\bf 129}, 177 (1983);
S.~Weinberg,
``Anthropic bound on the cosmological constant,''
Phys.\ Rev.\ Lett.\  {\bf 59}, 2607 (1987);
M.~Dine, ed., {\em String Theory in Four Dimensions},
Amsterdam: Elsevier Science Publishers (1988) and references therein.

\bibitem{msw}
H.~Martel, P.~R.~Shapiro, and S.~Weinberg,
``Likely values of the cosmological constant,''
Astrophys.\ J.\  {\bf 492}, 29 (1998)
[astro-ph/9701099].


}
\end{thebibliography}
\end{document}